\documentclass[11pt,a4paper]{article}
\usepackage{fullpage}
\newtheorem{theorem}{Theorem}

\newcommand{\polylog}{\ensuremath{\mathrm{polylog}}}
\newcommand{\occ}{\ensuremath{\mathrm{occ}}}

\begin{document}

\title{$r^*$-indexing}
\author{Travis Gagie}
\date{}
\maketitle

\begin{abstract}
Let $T [1..n]$ be a text over an alphabet of size $\sigma \in \polylog (n)$, let $r^*$ be the sum of the numbers of runs in the Burrows-Wheeler Transforms of $T$ and its reverse, and let $z$ be the number of phrases in the LZ77 parse of $T$.  We show how to store $T$ in $O (r^* \log (n / r^*) + z \log n)$ bits such that, given a pattern $P [1..m]$, we can report the locations of the $\occ$ occurrences of $P$ in $T$ in $O (m \log n + \occ \log^\epsilon n)$ time.  We can also report the position of the leftmost and rightmost occurrences of $P$ in $T$ in the same space and $O (m \log^\epsilon n)$ time.
\end{abstract}

Driven by pangenomics, indexed pattern matching in highly repetitive texts has become a hot topic in bioinformatics.  Gagie, Navarro and Prezza's $r$-index~\cite{GNP20} and its derivatives are probably the best-known options for this task.  Let $T [1..n]$ be a text over an alphabet of size $\sigma \in \polylog (n)$ and let $r$ be the number of runs in the Burrows-Wheeler Transform (BWT) of $T$.  Nishimoto and Tabei's~\cite{NT21} $r$-index-$f$ stores $T$ in $O (r \log n)$ bits such that, given a pattern $P [1..m]$, we can report the locations of the $\occ$ occurrences of $P$ in $T$ in $O (m + \occ)$ time.

Interestingly, if we want only to compute $\occ$ --- that is, to {\em count} the occurrences of $P$ but not {\em locate} them --- then we can use an $O (r \log (n / r)$-bit index.  This is the space we need for a run-length compressed BWT (RLBWT) for $T$ and some auxiliary data structures, with which we can compute in $O (m \log \log n)$ time the lexicographic interval of suffixes of $T$ that start with $P$, whose length is $\occ$.  We believe Brown et al.'s~\cite{BGMNS25} techniques can be used to reduce the counting time to $O (m)$ again without increasing the asymptotic space bound, but proving that formally is beyond the scope of this paper.

The difference between $O (r \log n)$ and $O (r \log (n / r))$ is often significant in practice, so researchers have tried to reduce the size of the suffix-array sample used to locate the occurrences as well.  Cobas, Gagie and Navarro's~\cite{CGN21,CGN??} {\it sr}-index uses a subsampled suffix-array sample and, depending on the data, can be significantly smaller than a standard $r$-index or the $r$-index-$f$.  As far as we know, however, every current implementation of the $r$-index or its derivatives uses $\Omega (r \log n)$ bits in the worst case to support locating.

There are other kinds of indexes for highly repetitive texts that approach locating differently.  For example, with K\"arkk\"ainen and Ukkonen's~\cite{KU96} LZ77-based index we keep a 2-dimensional range-reporting data structure for the $z \times z$ grid, where $z$ is the number of phrases in the LZ77 parse of $T$, with a point at $(x, y)$ if the co-lexicographically $x$th prefix ending at an LZ77 phrase boundary is immediately followed by the lexicographically $y$th suffix starting at an LZ77 phrase boundary.  We store the position of the phrase boundary corresponding to each point with the point, as satellite data.

Given $P$ we find, for $1 \leq i \leq z$, the co-lexicographic interval $[x_{i, 1}..x_{i, 2}]$ of the prefixes of $T$ ending at LZ77 phrase boundaries with $P [1..i]$ and the lexicographic interval $[y_{i, 1}..y_{i, 2}]$ of the suffixes of $T$ starting at LZ77 phrase boundaries with $P [i + 1..m]$, then find the points in the axis-aligned rectangle whose top left corner is $(x_{i, 1}, y_{i, 1})$ and whose bottom right corner is $(x_{i, 2}, y_{i, 2})$.  The satellite data for these points tells us the positions of phrase boundaries that split occurrences of $P$ into $P [1..i]$ and $P [i + 1..m]$.

We keep another 2-dimensional range-reporting data structure for the $n \times n$ grid with a point at $(x, y)$ if a phrase has source $T [x..y]$, with the starting position of that phrase stored as satellite data with the point.  If $s$ and $e$ are the starting and ending positions of an occurrence of $P$, then all the points to the left and above $(s, e)$ are sources that include $T [s..e]$ and whose phrases thus include copies of $P$.  Since every occurrence of $P$ either touches a phrase boundary or is completely included in a phrase whose source includes a copy of $P$, we can use these two range-reporting data structures to locate all the occurrences of $P$ in $T$.

K\"arkk\"ainen and Ukkonen's index was the beginning of a long line of research, culminating in Christiansen et al.'s~\cite{CEKNP20} $O (\gamma \log (n / \gamma))$-word index, where $\gamma$ is the size of the smallest string attractor~\cite{KP18} for $T$, that supports locating in $O (m + \occ \log^\epsilon n)$ time.  Converting Christiansen et al.'s space bound into bits we get $O (\gamma \log (n / \gamma) \log n)$, which does not seem directly comparable to $O (r \log (n / r))$~\cite{Nav21}.

There have already been attempts to combine the $r$-index with LZ77-indexes~\cite{FGHP14,FKP18,DAGHLMN24} but they required limiting the length of $P$ and had limited impact.  Our idea is to use an RLBWT for $T$ to find the lexicographic range of suffixes of $T$ starting $P [i + 1..m]$, and an RLBWT for $T$'s reverse to find the co-lexicographic range of prefixes of $T$ ending $P [1..i]$, for $1 \leq i \leq m$.  Letting $r^*$ denote the sum of the numbers of runs in the BWTs of $T$ and its reverse, these RLBWTs take $O (r^* \log (n / r^*))$ bits and querying them takes $O (m \log \log n)$ total time.

We keep one sparse bitvector with 1s marking the positions in the BWT of $T$ of characters occurring immediately to the right of an LZ77 phrase boundary in $T$, and another with 1s marking the positions in the BWT of $T$'s reverse of characters occurring immediately to the left of an LZ77 phrase boundary in $T$.  These take $O (z \log (n / z))$ bits and with them we can convert the ranges we find the RLBWTs into $[x_{1, 1}..x_{1, 2}], \ldots, [x_{m, 1}..x_{m, 2}]$ and $[y_{1, 1}..y_{1, 2}], \ldots, [y_{m, 1}..y_{m, 2}]$, again in $O (m \log \log n)$ time.

Since the range-reporting data structures each have $O (z)$ points and our queries to the second are only 2-sided, we can store them in $O (z)$ words --- so $O (z \log n)$ bits --- such that queries take a total of $O (m \log z + \occ \log^\epsilon n)$ time~\cite{Cha88,Lew13}.  Summing up and simplifying, we use $O (r^* \log (n / r^*) + z \log n)$ bits and $O (m \log n + \occ \log^\epsilon n)$ time.

\begin{theorem}
We can store $T$ in $O (r^* \log (n / r^*) + z \log n)$ bits such that when given $P$, we can report its locations in $T$ in $O (m \log n + \occ \log^\epsilon n)$ time.
\end{theorem}

For some applications, particularly taxonomic classification, we may be interested in finding the leftmost occurrence of $P$ in $T$~\cite{GKL22}.  In such cases, because the first occurrence of $P$ is guaranteed to touch an LZ77 phrase boundary, we can replace the first range-reporting data structure with an $O (z \log n)$-bit 2-dimensional range-minimum data structure with $O (\log^\epsilon n)$ query time~\cite{Nek21}, and discard the second range-reporting data structure.  Finding the rightmost occurrence is symmetric, on $T$'s reverse.

\begin{theorem}
We can store $T$ in $O (r^* \log (n / r^*) + z \log n)$ bits such that when given $P$, we can report its leftmost location in $T$ in $O (m \log^\epsilon n)$ time.
\end{theorem}

\end{document}